\documentstyle[a4,11pt]{article}
\newcommand{\al}{\alpha}
\newcommand{\be}{\beta}

\newcommand{\de}{\delta}
\newcommand{\ep}{\epsilon}
\newcommand{\ve}{\varepsilon}

\newcommand{\et}{\eta}

\newcommand{\ka}{\kappa}

\newcommand{\rh}{\rho}

\newcommand{\si}{\sigma}

\newcommand{\ps}{\psi}
\newcommand{\om}{\omega}
\newcommand{\Ga}{\Gamma}

\newcommand{\Si}{\Sigma}

\newcommand{\ti}{\times}

\begin{document}

%%%%%%%%%%%%%%%%%%%%%%%%%%%%%%%%%%%%%%%%%%%%%%%%%%%%%%%%%%%%%%%%%%%%%%%%
%Definition of signs for vector spaces
%%%%%%%%%%%%%%%%%%%%%%%%%%%%%%%%%%%%%%%%%%%%%%%%%%%%%%%%%%%%%%%%%%%%%%%%
\newfont{\sym}{msbm10 scaled 1200}
\newcommand{\RC}{\mbox{\sym C}} 
%complex numbers
\newcommand{\RR}{\mbox{\sym R}} 
%real numbers

%%%%%%%%%%%%%%%%%%%%%%%%%%%%%%%%%%%%%%%%%%%%%%%%%%%%%%%%
\sloppy
\pagestyle{headings}

%%%%%%%%%%%%%%%%%%%%%%%%%%%%%%%%%%%%%%%%%%%%%%%%%%%%%%%%
% titlepage
%%%%%%%%%%%%%%%%%%%%%%%%%%%%%%%%%%%%%%%%%%%%%%%%%%%%%%%%

\title{Non-closure of constraint algebra in N=1 supergravity}
\author{Matthias Wulf\thanks{M.Wulf@damtp.cam.ac.uk, permanent address: 
        Cognitive Systems Group (KOGS), Department of Computer Science,
University 
        of Hamburg, Vogt-K{\"o}lln-Str. 30, 22527 Hamburg, Germany,
        wulf@kogs.informatik.uni-hamburg.de}\\
Department of Applied Mathematics and Theoretical Physics \\
        Silver Street, Cambridge CB3 9EW, England }
\date{published in Int.J.Mod.Phys.D, Vol. 6, No. 1 (1997), 107-117,
  posted with kind permission of World Scientific Publishing Co.}
\maketitle
\abstract{The algebra of constraints arising in the canonical
          quantization of N=1 supergravity in four dimensions  is
investigated. Using the 
                holomorphic action, the structure functions of the 
        algebra are given and it is shown that the algebra 
                does not close formally for two chosen operator
orderings.}

%%%%%%%%%%%%%%%%%%%%%%%%%%%%%%%%%%%%%%%%%%%%%%%%%%%%%%%
\section{Introduction}
%%%%%%%%%%%%%%%%%%%%%%%%%%%%%%%%%%%%%%%%%%%%%%%%%%%%%%%

N=1 supergravity, the simplest supersymmetric extension of general
relativity, was first set up in \cite{FFN76} and \cite{DZ76}.  
Being nonrenormalizable but
finite up to second order in $\hbar$ in the perturbative expansion,
finiteness at all orders is unlikely for
the unbounded case \cite{DKS77} but still under debate in presence of
boundaries \cite{D'EXX}. 

What makes locally supersymmetric theories interesting in the
canonical approach  is the fact that the commutator of two
supersymmetry transformations gives a general coordinate
transformation. Hence the physical states, i.e. those state
functionals that are annihilated by the quantized constraints
corresponding to those transformations, are easier to find, since one
only has to look for solutions to the supersymmetry constraints to
find  states that are also invariant under general coordinate
transformations. 

In the framework of canonical quantization of theories with constraints
\cite{Dir65} 
a crucial aspect is that the quantized constraints are required to form 
an algebra in order for the quantum theory to be consistent. This means
that 
the commutator of two constraints should give an expression of the form 
{\it structure function $\ti$ constraint} with the constraint operator 
standing on the right so that the commutator of two constraints that 
annihilate a physical state also annihilates this state. Although the
classical 
constraint algebra has fully been given in \cite{Tei77}, a check of the
more 
involved terms of the quantized algebra is necessary. 

The starting point for the canonical quantization is the action 
of N=1 supergravity. It is chosen to work with the holomorphic action 
\cite{Jac88},  
the conventions being according to \cite{D'E84}, as set out in the
appendix.
\begin{eqnarray}   
  \tilde{I}[e, \om , \ps , \ps^{\ti} ] & = & - \int d^4 x \;   \ve^{\mu
\nu \rh \si}
        \; \Big( \ps^{\ti A'}{}_{\mu} \; e_{AA' \nu} \; D_{\rh}
\ps^{A}{}_{\si} 
        \nonumber \\
  & & {} + \frac{i}{\ka^2} \; e_{AA' \rh} \; e^{A'}{}_{B \si} \; 
        [\partial_{\mu} \om^{AB}{}_{\nu} + \om^{A}{}_{C \mu} \;
\om^{CB}{}_{\nu}
        ] \Big)
\label{action}
\end{eqnarray}
Here $\ve^{\mu \nu \rh \si}$ is the L\'{e}vi-Civita tensor density
with $- \ve^{0123} = \ve_{0123} =1$. $\ps^{A}{}_{\si}$, $A = 0,1$, are
the components of a spinor-valued one-form describing the spin-3/2
degrees of freedom, and hence are Grassmann valued. The spinor $\ps^{\ti
A'}{}_{\si}$, $A' = 0', 1'$, 
corresponding to the complex conjugate of $\ps^{A}{}_{\si}$ 
in the real theory , is considered to be independent
since the complex conjugate of a holomorphic function is not
holomorphic. $e^{AA'}{}_{\mu}$ are
spinor-valued tetrads standing for the gravitational or spin-2 degrees
of freedom and are taken to be invertible. Indices from the middle of
the greek alphabet are
spacetime indices while those from the middle of the lower
case latin alphabet are spatial indices. 
The constant $\ka^2$ takes the value $8 \pi$.
Covariant
derivatives appear only antisymmetrized in the spacetime indices and
are given by
\begin{eqnarray*}   
        D_{[ \mu} \ps^{A}{}_{\nu ]}  & = & \partial_{[ \mu}
\ps^{A}{}_{\nu ]} 
                + \om^{A}{}_{B [ \mu} \; \ps^{B}{}_{\nu ]} 
\end{eqnarray*}
hence the symmetric spacetime connection $\Ga^{\mu}_{\nu \rh}$ never
appears unlike the spin connection $\om^{AB}{}_{\mu}$ which will be
abbreviated as $\om$ throughout the text. $\om$ is treated according
to the {\it 1.5 order method} \cite{vN81}, i.e. one takes it to be an
independent variable first, solves its (nonpropagative) equation of
motion 
leading to a solution for $\om$ as a function of the tetrad $e$ and
the spinors $\ps$ and $\ps^{\ti}$ which then is inserted back into the
action. However, being a solution of its own equation of motion,
i.e. $\de \tilde{I} / \de \om = 0$, it is not necessary to
differentiate the $\om$'s when it comes to differentiate the action by
the other fields. 

The variables $e$, $\ps$ and $\ps^{\ti}$ have to obey reality
conditions, given below,  to make the theory equivalent to the real
theory. 
 The
equations of motion arising from this action are known to be the same
as those of the real theory after insertion of the reality conditions
\cite{Mat94b,Wul94}. Due to the complexification of the theory
the Lorentz algebra splits into two factors, one with $\om$ as a gauge
field, the other with $\bar{\om}$. Since the latter does not appear in
the action, the two factors differ considerably \cite{Mat94b}. 
The theory is symmetric under general coordinate transformations, the
variation 
of the fields given by their Lie derivative,
as well as under left-handed local Lorentz transformations
\begin{displaymath}  
   \de e^{AA'}{}_{\mu}  =  N^{A}{}_{B} \; e^{BA'}{}_{\mu} \qquad
   \de \ps^{A}{}_{\mu}  =  N^{A}{}_{B} \; \ps^{B}{}_{\mu} \qquad
   \de \ps^{\ti A'}{}_{\mu}  =  0
\end{displaymath}
with a parameter $N^{AB} = N^{BA}$ and correspondingly right-handed
Lorentz transformations
\begin{displaymath}  
   \de e^{AA'}{}_{\mu}  =  \bar{N}^{A'}{}_{B'} \; e^{AB'}{}_{\mu} \qquad
   \de \ps^{A}{}_{\mu}  =  0 \qquad
   \de \ps^{\ti A'}{}_{\mu}  =  \bar{N}^{A'}{}_{B'} \; 
        \ps^{\ti B'}{}_{\mu}
\end{displaymath}
with a parameter $\bar{N}^{A'B'} = \bar{N}^{B'A'}$. Finally there is
the symmetry under left-handed supersymmetry transformations 
\begin{displaymath}  
   \de e^{AA'}{}_{\mu}  =   - \frac{i \ka^2}{2} \; \ep^{A} \; \ps^{\ti
A'}{}_{\mu} 
        \qquad
     \de \ps^{A}{}_{\mu}  =   D_{\mu} \ep^{A} \qquad
  \de \ps^{\ti A'}{}_{\mu}   =  0
\end{displaymath}
with Grassmann valued parameters $\ep^{A}$. The transformation 
of $\ps^{\ti A'}{}_{\mu}$ under
right-handed supersymmetry is, since there is no $\bar{\om}$ to give
$D_{\mu} \bar{\ep}^{A'}$, more complicated \cite{Mat94b}. However,
after inserting the reality conditions into the transformations one gets
\begin{displaymath}  
    \de e^{AA'}{}_{\mu}  =   - \frac{i \ka^2}{2} \; \bar{\ep}^{A'} \;
\ps^{A}{}_{\mu} 
        \qquad
     \de \ps^{A}{}_{\mu}  =   0 \qquad
  \de \bar{\ps}^{A'}{}_{\mu}   =  D_{\mu} \bar{\ep}^{A'}
\end{displaymath}
Formulating the theory in a canonical way lets one find the constraints
which generate 
these transformations.

%%%%%%%%%%%%%%%%%%%%%%%%%%%%%%%%%%%%%%%%%%%%%%%%%%%%%%%%%%%%%%%%%%%%%%%%%%
\section{Canonical Formulation}
%%%%%%%%%%%%%%%%%%%%%%%%%%%%%%%%%%%%%%%%%%%%%%%%%%%%%%%%%%%%%%%%%%%%%%%%%%

To get the canonical formulation of supergravity spacetime is split
into space and time according to \cite{MTW73} limiting the topology of
spacetime to be $\Si \ti \RR$, where $\Si$ is a spatial
hypersurface. The ''time'' associated with $\RR$ is just a parameter and 
to be distinguished from a - difficult to define - physical time
\cite{Mat94b}. 
An effect of this spacetime split is that the invariance under general
coordinate transformations splits into one under translations of the
time parameter and one under spatial diffeomorphisms.

The spinor equivalent of the outward normal vector to the hypersurface
$\Si$ 
is given by 
$n^{AA'}$ with
\begin{displaymath}
  n_{AA'} \; n^{AA'} = 1 \qquad \mbox{and} \qquad n_{AA'} \;
  e^{AA'}{}_{i} = 0
\end{displaymath}
which is a function of the spatial components of $e$ (see
apendix). The time component can be written as 
\begin{displaymath}
  e^{AA'}{}_{0} = N \; n^{AA'} + N^{i} \; e^{AA'}{}_{i}
\end{displaymath}
where $N$ is the {\it Lapse} and $N^{i}$ the {\it Shift} functions
\cite{MTW73}. 
Calculating the momenta from (\ref{action}) one has to be aware of the
Grassmann valuedness of the spin-3/2 variables, hence 
anticommute these variables to the left before performing functional
differentiation on them. The momenta of the theory are
\begin{eqnarray}   
  \pi_{A}{}^{j} & := & \frac{\de \tilde{I}}{\de \dot{\ps}^{A}{}_{j}} = 
        - \ve^{ilj} \; \ps^{\ti A'}{}_{i} \; e_{AA' l}
  \label{3.5} \\
  p_{AA'}{}^{j} & := & \frac{\de \tilde{I}}{\de \dot{e}^{AA'}{}_{j}} = 
        - \frac{2i}{\ka^2} \; \ve^{jki} \; e_{BA'k} \; \om^{B}{}_{Ai}
  \label{3.6} 
\end{eqnarray}
Due to the 1.5 order method $\om$ is not treated as a canonical
variable hence it has no corresponding momentum. One clear advantage
of working with the holomorphic action can be seen looking at 
(\ref{3.5}) which involves $\ps^{\ti}$. In the real theory there is a
similar expression for the momentum of $\bar{\ps}$ \cite{D'E84}, so
the four variables $\ps$, $\pi$, $\bar{\ps}$ and $\bar{\pi}$ are not
independent and give rise to second class constraints whose treatment
needs the construction of Dirac brackets \cite{Dir65} whereas here one
can 
treat $\ps$ and $\pi$ as independent variables. 

Choosing $e$, $\ps$ and $p$, $\pi$ from (\ref{3.5}) and (\ref{3.6}) as 
canonical variables, the next step is to define Poisson brackets.
Holomorphic Poisson brackets for holomorphic functionals $F$ and $G$
of the canonical variables are defined by 
\pagebreak[4]
\begin{eqnarray*}    
  \{F,G \} & := & \int d^3 u \; \left( \frac{\de G}{\de
p_{AA'}{}^{i}(u)} 
        \frac{\de F}{\de e^{AA'}{}_{i}(u)}      \;  
        - \frac{\de G}{\de e^{AA'}{}_{i}(u)} \; \frac{\de F}{\de
p_{AA'}{}^{i}(u)}
        \right) \\ 
        & & {} - \left(
        \frac{\de G}{\de \pi^{A}{}_{i}(u)} \; \frac{\de F}{\de
\ps^{A}{}_{i}(u)}
        + \frac{\de G}{\de \ps^{A}{}_{i}(u)} \; 
        \frac{\de F}{\de \pi_{A}{}^{i}(u)} \right)      
\end{eqnarray*}
being symmetric for the fermionic derivatives and obeying the rules set 
up in \cite{Cas76}. With  (\ref{3.5}) and (\ref{3.6}) follows
\begin{eqnarray}  
  \{ \pi_{B}{}^{j}(x) , \ps^{A}{}_{i}(y) \} & = & - \ep_{B}{}^{A} \;
\de_{i}{}^{j} \; 
        \de (x,y) \label{Poiss1} \\
  \{ p_{BB'}{}^{j} , e^{AA'}{}_{i} \} & = & - \ep_{B}{}^{A} \;
\ep_{B'}{}^{A'}
        \; \de_{i}{}^{j} \; \de (x,y) \label{Poiss2}
\end{eqnarray}
which are the only nonvanishing brackets.

Before coming to the constraints, it is useful to discuss the reality 
conditions.
The reality conditions are given by
\begin{eqnarray*}    
  {R_1}^{AA'}{}_{i} & := &  e^{AA'}{}_{i} \\
  {R_2}^{AA'k} & := &  i \;  \ve^{ijk} \; \ps^{A}{}_{i} \; \ps^{\ti
A'}{}_{j} \\
  {R_3}_{AA'}{}^{j} & := & p_{AA'}{}^{j} + \frac{i}{\ka^2} \; \ve^{ijk}
\; 
        \partial_i e_{AA'k} - \frac{1}{2} \; \ve^{ijk} \;
\ps^{\ti}{}_{A'i} \; \ps_{Ak}
\end{eqnarray*}
The first two conditions state the reality of $e$ and the fact that 
$\ps^{\ti}$ is the complex conjugate of $\ps$ in the real theory. 
The third reality condition arises from claiming that $p + p^{\ti}$ 
should be real, $p^{\ti}$ being a holomorphic function corresponding to
$\bar{p}$ 
after insertion of the first two reality conditions. However, $p$ itself 
is not required to be real \cite{Jac88}. 
The resulting second class constraints ${\rm Im}(R_a) \approx 0$, 
$\approx 0$ meaning ''weakly zero'' \cite{MTW73}, cause
no problems as the Dirac brackets that follow from them are equal to
the holomorphic Poisson brackets \cite{Mat94b,Wul94}. Hence for
each nonholomorphic field $\bar{F}$ a holomorphic field $F^{\ti}$ can
be found, being equal to $\bar{F}$ modulo the reality conditions, and
can be used instead, since
\begin{displaymath}
 \{ G, \bar{F} \}_{\ast} = \{ G, F^{\ti} \}_{\ast} = \{ G, F^{\ti} \}
\end{displaymath}
where $\{ \quad , \quad \}_{\ast}$ are the Dirac brackets with respect
to 
${\rm Im}(R_i) \approx 0$.
${R_1}^{AA'}{}_{i}$ and the projections ${R_2}_{[BB'}{}^{(s} \;
e_{C]}{}^{A'p)}$, 
$( \ldots )$ denoting symmetrization in the indices, 
form a set of 18 commuting reality conditions, meaning
that there is a real configuration space, described by those variables
whose reality is enforced by these 18 conditions. In the quantized
theory the reality conditions will become exact operator identities
that restrict the possible scalar product of physical states. 

The constraints arise as follows. 
A primary constraint follows from (\ref{3.6})
\begin{equation}    
  {\cal J}^{\ti}_{A'B'} := - e^{A}{}_{(A'j} \; p_{AB')}{}^{j} \approx 0
  \label{3.7} 
\end{equation}
The variation of the canonical variables as given by 
$\de\chi = \{ \chi, \int d^3y \, \bar{N}^{A'B'} \, {\cal J}^{\ti}_{A'B'}
\}$ 
with the parameter $\bar{N}^{A'B'} = \bar{N}^{B'A'}$ are 
\begin{equation}    
  \de e^{DD'}{}_{s} = \bar{N}_{A'}{}^{D'} \, e^{DA'}{}_{s} \quad 
  \de p_{DD'}{}^{s} = - \bar{N}_{D'}{}^{A'} \, p_{DA'}{}^{s} \quad
  \de \ps^{D}{}_{s} = 0 \quad
  \de \pi_{D}{}^{s} = 0
\label{rhTrans}
\end{equation}
thus identifying ${\cal J}^{\ti}$ as the generator of right-handed 
Lorentz transformations. 
To find the generator of left-handed Lorentz transformations, 
which in the real 
theory is the complex conjugate of ${\cal J}^{\ti}$, one takes the
complex 
conjugate of (\ref{3.6}), uses the torsion equation of the real theory
to get 
a holomorphic function $\bar{\om}$ and finally uses the reality
conditions 
to replace the remaining nonholomorphic 
variables by holomorphic ones. This leads to
\begin{equation}    
  \bar{p}_{AA'}{}^{j} = p_{AA'}{}^{j} + \frac{2i}{\ka^2} \; \ve^{ijk} \; 
        \partial_{i} e_{AA'k} - \ve^{ijk} \; \bar{\ps}_{A'i} \; \ps_{Ak}
 \label{pquer} 
\end{equation}
With this, one finds the generator of left-handed Lorentz
transformations.
\begin{eqnarray}   
  {\cal J}_{AB} & = & - e^{A'}{}_{(Ak} \; p_{B)A'}{}^{k} +
\frac{i}{\ka^2} \; 
        \partial_i (\ve^{ikl} \; e_{(AA'k} \; e^{A'}{}_{B)l}) 
        -   \ps_{(Bl} \; \pi_{A)}{}^{l}  \label{3.13} \\
  & = & \frac{i}{\ka^2} \; D_i (\ve^{ikl} \; e_{(AA'k} \;
e^{A'}{}_{B)l}) 
        -  \; \ps_{(Bl} \; \pi_{A)}{}^{l} \approx 0
  \nonumber 
\end{eqnarray}
The
secondary constraints are given by
\begin{eqnarray}   
  {\cal S}_{A}  : =  \frac{\de L}{\de \ps^{A}{}_{0}} & = & D_i
\pi_{A}{}^{i}
        \approx 0 \label{3.14} \\
  {\cal S}^{\ti}_{A'}  :=  -  \frac{\de L}{\de \ps^{\ti A'}{}_{0}} & = &
        - \ve^{ijk} \; e_{AA'i} \; D_j \ps^{A}{}_{k}
        \approx 0 \label{3.15} \\
  {\cal H}_{AA'}  :=  \frac{\de L}{\de e^{AA'}{}_{0}} & = & 
        \frac{2i}{\ka^2} \; \ve^{ijk} \; e_{BA'k} \; [\partial_i
\om^{B}{}_{Aj}
        + \om_{ACi} \; \om^{CB}{}_{j} ] \nonumber \\
  & & {} - \ve^{ijk} \; \ps^{\ti}{}_{A'i} \; D_j \ps_{Ak}
        \approx 0 \label{3.16}  
\end{eqnarray} 
where ${\cal S}_A$ is the generator of left-handed, ${\cal
S}^{\ti}_{A'}$ that of right-handed supersymmetry transformations and
${\cal H}_{AA'}$ the combined generator of time translations
(Wheeler-deWitt generator) and of spatial diffeomorphisms plus
Lorentz and supersymmetry
transformations \cite{Wul94}. 

To express the canonical Hamiltonian density and hence the secondary 
constraints in terms of the canonical variables, 
 it is necessary to invert
(\ref{3.5}) and (\ref{3.6}). This inversion takes place on the 
hypersurface in phase space given by the vanishing of the primary 
constraint ${\cal J}^{\ti}$, the surface on which 
the canonical Hamiltonian is defined, and leads to 
\begin{eqnarray}    
  \ps^{\ti B'}{}_{i} & = &  \pi_{A}{}^{j} \; D^{AB'}_{ji}
         \label{3.8} \\
  \om^{AB}{}_{i}  & = & - \frac{i \ka^2}{2} \;   p_{A'}{}^{Bj} \;
D^{AA'}_{ij}
\label{3.11} 
\end{eqnarray}
with
\begin{eqnarray}   
  D^{AA'}_{jk} & := & - \frac{2i}{\sqrt{h}} \; e^{AB'}{}_{k} \; e_{BB'j}
\; 
        n^{BA'} \nonumber \\
  & = & \ve_{jkp} \; e^{AA'p}  + \frac{i}{\sqrt{h}} \; h_{jk} \; n^{AA'}
  \label{3.9},
\end{eqnarray}
because of
\begin{equation}    
  D^{CE'}_{rj} \; \ve^{lrm} \; e_{CC'm} = \ep_{C'}{}^{E'} \;
\de_{j}{}^{l} 
  \qquad
  D^{CE'}_{rj} \; \ve^{jlm} \; e_{DE'm} = \ep_{D}{}^{C} \; \de_{r}{}^{l}
\label{3.10} 
\end{equation} 
On this hypersurface, the part of the rhs of (\ref{3.11}) 
that is antisymmetric in $A$ and $B$ vanishes, yielding an expression
for
$\om$ with the correct number of degrees of freedom. 
 ${\cal J}$ and ${\cal
J}^{\ti}$ are multiplied by Langrangian multipliers $\om^{AB}{}_{0}$
and $\bar{\om}^{A'B'}{}_{0}$ and added to the canonical Hamiltonian
density to give the total Hamiltonian density
\begin{displaymath}    
  {\cal H}_t = - e^{AA'}{}_{0} \; {\cal H}_{AA'} - \om^{AB}{}_{0} \; 
        {\cal J}_{AB} - \ps^{A}{}_{0} \;  {\cal S}_{A} - 
        {\cal S}^{\ti}_{A'} \; \ps^{\ti A'}{}_{0}
        - \bar{\om}^{A'B'}{}_{0} \; \bar{\cal J}_{A'B'} 
\end{displaymath}
which is the typical picture in reparametrization-invariant theories: 
The total Hamiltonian vanishes weakly. 
The secondary 
constraints can now all be given as functions of the canonical variables
\begin{eqnarray}   
  {\cal S}_{A} & = & \partial_i \pi_{A}{}^{i} + \frac{i \ka^2}{2} \; 
        p_{AA'}{}^{j} \; D^{BA'}_{ij} \;  \pi_{B}{}^{i}
  \label{Sc} \\
  {\cal S}^{\ti}_{A'} & = & - \ve^{ijk} \; e_{AA'i} \; \partial_j
\ps^{A}{}_{k} 
        - \frac{i \ka^2}{2} \; p_{BA'}{}^{k} \; \ps^{B}{}_{k}
   \label{Squerc} \\
   {\cal H}_{AA'} & = & \partial_l p_{AA'}{}^{l} - \ve^{ijk} \; 
        \; p_{AC'}{}^{l} \; D^{BC'}_{jl} \; \partial_i
        e_{BA'k}   \;
         + \frac{i \ka^2}{2} \;  p_{AC'}{}^{l} \; D^{CC'}_{il} \; 
        p_{CA'}{}^{i} \nonumber \\
  & & {} + \frac{i \ka^2}{2} \; \ve^{ijk} \;  p_{AC'}{}^{m} \;
D^{B}{}_{A'li} \; 
        D^{CC'}_{jm}  \; \ps_{Ck} \; \pi_{B}{}^{l} \nonumber \\
  & & {} + \ve^{ijk} \; D^{B}{}_{A'li} \; \partial_j ( \ps_{Ak}) \; 
        \pi_{B}{}^{l}   
   \label{Hc} 
\end{eqnarray}

Now that the canonical variables and 
constraints are known it is possible to proceed to quantize the theory.

%%%%%%%%%%%%%%%%%%%%%%%%%%%%%%%%%%%%%%%%%%%%%%%%%%%%%%%%%%%%%%%%%%%%%%%
\section{Constraint Algebra}
%%%%%%%%%%%%%%%%%%%%%%%%%%%%%%%%%%%%%%%%%%%%%%%%%%%%%%%%%%%%%%%%%%%%%%%

To quantize the theory canonically, one has to find operators
corresponding to the canonical variables fulfilling the following
quantization prescription for even variables $E$ and odd variables $O$ 
 \begin{eqnarray*}  
  [ \hat{E}_1 , \hat{E}_2 ]   =  i \hbar \; \widehat{ \{ E_1 , E_2 \}} 
\quad 
  [ \hat{O} , \hat{E} ]  =  i \hbar \; \widehat{\{ O , E \}} \quad
  [ \hat{O}_1 , \hat{O}_2 ]_{+}  =  i \hbar \; \widehat{\{ O_1 , O_2 \}}
\end{eqnarray*}
where $[ \quad , \quad ]_{+}$ stands for the anticommutator.
It is not necessary to consider a specific representation of the
operators that 
correspond to the canonical variables because for the algebra of
constraints one only 
needs the commutation relations of those operators that are 
given by  
(\ref{Poiss1}) and (\ref{Poiss2}) multiplied by $i \hbar$. 
A representation giving the correct form 
of the Lorentz generators is given in \cite{Mat94a,Wul94}.

Using the equations (\ref{Sc}) to (\ref{Hc}) as the quantum constraints
with 
the given operator ordering and employing 
\begin{equation}    
  [ p_{AA'}{}^{i}(x) , D^{BB'}_{jk}(y) ] = i \hbar \; \ve^{rsi} \; 
        D^{B}{}_{A'jr} \; D^{B'}{}_{Ask} \, \de(x,y)
  \label{4.25}
\end{equation}
(which follows from (\ref{3.9})) one gets the well-known results
\begin{eqnarray*}
  [ {\cal S}^{\ti}_{A'}(x) , {\cal S}^{\ti}_{B'}(y) ]_{+}  =  0 
  & \qquad  &
  [ {\cal S}_{A}(x) ,  {\cal S}_{B}(y) ]_{+} = 0
  \\
    {} [   {\cal S}_{A}(x) , {\cal S}^{\ti}_{A'}(y) ]_{+} & = &
  - \frac{\hbar \ka^2}{2} \; {\cal H}_{AA'} \; \de (x,y)
\end{eqnarray*}
To allow for partial integration in these calculations the constraints
have 
been contracted with Grassmann valued transformation parameters and
integrated 
over $x$ and $y$. 
Also, the partial derivative of the square of the delta function is
taken to be zero. 
It is assumed that one can find regularized operators for the theory
that fulfill 
this requirement. 
The calculation of 
$[{\cal S}^{\ti}_{A'} , {\cal H}_{BB'} ]$ can be performed in the same 
straightforward manner using (\ref{6.2}) and (\ref{3.13}) yielding
\begin{eqnarray*}    
  \lefteqn{[{\cal S}^{\ti}_{A'}(x) , {\cal H}_{BB'}(y) ]  =  } \\
         & &  i \hbar \ka^2 \; \ve^{lmn} \; \ep_{A'B'} \; 
         ( \partial_m \ps_{Bn} + \frac{i \ka^2}{2} \; p_{BD'}{}^{q} \; 
        D^{ED'}_{mq} \; \ps_{En} ) \; \ti \\
 & & \frac{1}{\sqrt{h}} \; n^{C}{}_{G'} \; e^{GG'}{}_{l} \; {\cal
J}_{CG} \; 
        \de(x,y) \\
 & = &  i \hbar \ka^2 \; \ve^{lmn} \; \ep_{A'B'} \; 
        D_m \ps_{Bn} \; \frac{1}{\sqrt{h}} \; n^{C}{}_{G'} \;
e^{GG'}{}_{l} \; 
        {\cal J}_{CG} \; \de(x,y) \; 
\end{eqnarray*}
where in the last line the correspondence between (\ref{3.14}) and
(\ref{Sc}) 
with the chosen operator ordering was used to define the ordering of an
operator 
version of $\om$. 
Since the result is a constraint 
times a structure function appearing on the left hand side, this
commutator 
shows no sign of non-closure of the algebra of constraints. 

Working out $[{\cal S}_{A} , {\cal H}_{BB'}]$ by the same methods,
making 
use of (\ref{4.27}) and 
\begin{displaymath}    
  D^{CC'}_{sj} \; p_{CC'}{}^{j} = - \frac{2 i}{\sqrt{h}} \; n^{GC'} \; 
        e^{G'}{}_{Cs} \; {\cal J}^{\ti}_{G'C'}
\end{displaymath}
leads to 
\begin{eqnarray}  
  \lefteqn{[{\cal S}_{A}(x) , {\cal H}_{BB'}(y)]  = 
        \frac{i \hbar \ka^2}{\sqrt{h}} \; \ve{lmn} \; \ve_{AB} \;
n^{GC'} \; 
        e^{G'}{}_{Gn} \; {\cal J}^{\ti}_{G'C'} \; \ti} \nonumber \\
  & & \Big( \partial_m \ps^{\ti}{}_{B'l} + \frac{i \ka^2}{2} \;
D^{CC'}_{im} \; 
        p_{CB'}{}^{i} \; \ps^{\ti}{}_{C'l} - \ve^{sqr} \; D^{DC'}_{qm}
\; 
        \partial_s e_{DB'r} \; \ps^{\ti}{}_{C'l} \nonumber \\
  & & + \frac{i \ka^2}{2} \; \ve^{rsq} \; D^{ED'}_{sm} \; \ps_{Eq} \; 
        \ps^{\ti}{}_{B'r} \; \ps^{\ti}{}_{C'l} \Big) \label{SH1}
\end{eqnarray}
where the terms in brackets can be interpreted as $D_m
\ps^{\ti}{}_{B'l}$ when 
choosing a holomorphic $\bar{\om}$ to have the above operator ordering.
Note that 
this ordering differs from that of $\om$ (\ref{3.11}) since ${\cal S}_A$ 
(\ref{3.14}) is used in its left-ordered form (\ref{Sc}). Commuting 
${\cal J}^{\ti}$ through to the right in (\ref{SH1}) using
(\ref{rhTrans}) 
gives rise to the divergent expression
\begin{eqnarray*}  
 \lefteqn{[{\cal S}_{A}(x) , {\cal H}_{BB'}(y)]_{divergent} = } \\
  & = & - \frac{\hbar \ka^2}{\sqrt{h}} \; \de(0) \; \ve^{lmn} \;
\ep_{AB} \; 
        n^{DD'} \; e_{DB'n} \; D_m \ps^{\ti}{}_{D'l} \; \de(x,y)
\end{eqnarray*}
Writing $D_m \ps^{\ti}{}_{D'l}$ as $(D_m D^{F}{}_{D'ql}) \pi_{F}{}^{q}$ 
plus $D^{F}{}_{D'ql} D_m \pi_{F}{}^{q}$ by properly introducing $\om$
and using
(\ref{3.10}) one gets one term involving ${\cal S}_D$ and one involving 
$D_m e^{D'}{}_{Dn}$. However, since in the latter expression $\bar{\om}$ 
appears right-ordered with respect to $p$ whereas $\om$ is left-ordered,
it 
leads to 
\begin{displaymath}    
  \ve^{lmn} \; D_m e^{D'}{}_{Dn} = - \frac{i \ka^2}{2} \; \ve^{lmn} \; 
        ( \ps_{Dm} \; \ps^{\ti D'}{}_{n} - D^{D'}{}_{Dmn} \; \de(0) )
\end{displaymath}
Introducing this into the above equation and using (\ref{anticom})
finally 
yields
\begin{eqnarray*}    
  \lefteqn{[{\cal S}_{A}(x) , {\cal H}_{BB'}(y)]_{divergent} = } \\
  & = & \frac{\hbar \ka^2}{\sqrt{h}} \; n^{D}{}_{B'} \; {\cal S}_{D} \;
\de(0) 
        \; \de(x,y)
        + \frac{\hbar \ka^4}{2 h} \; e^{F}{}_{B'k} \; \pi_{F}{}^{k} 
        \; (\de(0))^2 \; \de(x,y)
\end{eqnarray*}
The first term does not lead to difficulties, 
since it involves a constraint sitting on the right-hand 
side. The second term, however, clearly leads to 
non-closure of the algebra of these operators.   

Choosing the right-ordered version ${\cal S}^{\cal R}_{A}$
\begin{displaymath}    
  {\cal S}^{\cal R}_{A} = \partial_i \pi_{A}{}^{i} + \frac{i \ka^2}{2}
\; 
        D^{BB'}_{ij} \; p_{AB'}{}^{j} \; \pi_{B}{}^{i}
\end{displaymath}
leads, via $[{\cal S}^{\cal R}_{A}(x), {\cal S}^{\ti}_{A'}(y)] 
= - \frac{\hbar \ka^2}{2} \, {\cal H}^{\cal R}_{AA'} \, \de(x,y)$ to the 
expression
\begin{eqnarray*}  
  {\cal H}^{\cal R}_{AA'} & = & \partial_l p_{AA'}{}^{l} - \ve^{ijk} \; 
        \; D^{BC'}_{jl} \; p_{AC'}{}^{l} \;  \partial_i
        e_{BA'k}   \;
         + \frac{i \ka^2}{2} \;   D^{CC'}_{il} \; p_{AC'}{}^{l} \;
        p_{CA'}{}^{i}  \\
  & & {} + \frac{i \ka^2}{2} \; \ve^{ijk} \;   D^{B}{}_{A'li} \; 
        D^{CC'}_{jm}  \; p_{AC'}{}^{m} \; \ps_{Ck} \; \pi_{B}{}^{l}  \\
  & & {} + \ve^{ijk} \; D^{B}{}_{A'li} \; \partial_j ( \ps_{Ak}) \; 
        \pi_{B}{}^{l}   
\end{eqnarray*}
If one calculates the commutator $[ {\cal S}^{\ti}_{A'} , 
{\cal H}^{\cal R}_{BB'}]$, one arrives at
\begin{eqnarray*}  
  [ {\cal S}^{\ti}_{A'}(x) , {\cal H}^{\cal R}_{BB'}(y)] & = & 
        \frac{i \hbar \ka^2}{\sqrt{h}} \; \ep_{A'B'} \; \ve^{lmn} \;
n^{C}{}_{G'} \; 
        e^{GG'}{}_{l} \, \ti \\
 & & \Big( (\partial_m \ps_{Bn}) {\cal J}_{CG} + \frac{i \ka^2}{2} \; 
        D^{ED'}_{mq} \; \ps_{En} \; {\cal J}_{CG} \; p_{BD'}{}^{q} \Big)
\end{eqnarray*}
Commuting $p$ with ${\cal J}$ leads to divergent terms of the form
\begin{eqnarray*}  
  \lefteqn{[ {\cal S}^{\ti}_{A'}(x) , {\cal H}^{\cal R}_{BB'}(y)
        ]_{divergent}  = }\\
 & = & \frac{2i}{\sqrt{h}} \, \de(0) \; \ve^{lmn} \; n_{BG'} \;
e^{EG'}{}_{l} 
        \; n^{FD'} \; e^{C'}{}_{Fm} \;  \ps_{En} \; {\cal
J}^{\ti}_{D'C'} \\
 & & {} + i \hbar \, \de(0) \; n_{BG'} \; \ps_{En} \; ( p^{EG'n} \; 
        + \frac{2i}{\ka^2} \; \ve^{lmn} \; e^{GG'}_{l} \; D^{ED'}_{mq}
\; 
        \ve^{ijq} \; \partial_j e_{GD'i})
\end{eqnarray*}
Again, divergent terms without a constraint in the right-hand position
arise 
so that the algebra of the right-handed operators does not close either.
Apart from the closure issue, these simple and straightforward
calculations 
also gave the structure functions of the classical algebra.

%%%%%%%%%%%%%%%%%%%%%%%%%%%%%%%%%%%%%%%%%%%%%%%%%%%%%%%%%%
\section{Discussion and Acknowledgements}
%%%%%%%%%%%%%%%%%%%%%%%%%%%%%%%%%%%%%%%%%%%%%%%%%%%%%%%%%%

The above calculations show that using the holomorphic formulation 
of N=1 supergravity one can see that there 
is no formal closure of the constraint algebra
for the two orderings chosen. This means that - in the sense 
of Dirac \cite{Dir65} - the canonical quantization has 
failed since it leads to inconsistencies. A different viewpoint would be
to take e.g. 
$[ {\cal S}_A , {\cal H}_{BB'}]$ on as a new constraint. However, it  
 would still be necessary to verify the  closure of the entire algebra. 
In any case it substantially reduces the set of physical states. Whether
this 
still remains a meaningful theory is a topic for further investigation
 as well as the question whether the non-closure holds for all possible
operator 
orderings. 

It has to be kept in mind that physically meaningful results concerning
the 
 algebra can only be derived using regularized operators, since in the
formal 
 calculations delta function identities are used \cite{TW87,FJ88}. 
 The methods and results presented here hence are paving the way for a
more 
 involved regulated calculation. They also serve 
 as a further demonstration showing the usefulness of the holomorphic
formulation of 
 supergravity: 
 The expressions for the constraints and hence the calculation of the
algebra 
 become relatively simple as compared to the real theory \cite{D'E84}.
 Also, giving the structure functions of the quantum algebra
explicitely, 
 those of the classical algebra, as given in \cite{Tei77}, are found 
 as well.

\bigskip
\noindent
The author would like to thank Peter D'Eath, Hermann Nicolai and
especially 
Hans-J{\"u}rgen Matschull for many useful hints and interesting
discussions.

%%%%%%%%%%%%%%%%%%%%%%%%%%%%%%%%%%%%%%%%%%%%%%%%%%%%%%%%
%%%%%%%%%%%%%%%%%%%%%%%%%%%%%%%%%%%%%%%%%%%%%%%%%%%%%%%%
% appendix
%%%%%%%%%%%%%%%%%%%%%%%%%%%%%%%%%%%%%%%%%%%%%%%%%%%%%%%%

%%%%%%%%%%%%%%%%%%%%%%%%%%%%%%%%%%%%%%%%%%%%%%%%%%%%%%%%
\section{Appendix}

Throughout the text spinor-valued tetrads $e^{AA'}{}_{\mu}$ are used to 
describe the gravitational degrees of freedom. Spinor indices take the 
values $0$ and $1$ or, respectively, $0'$ and $1'$. The indices $\mu$,
$\nu$, 
$\rh$ $\ldots$ are spacetime indices taking values from $0$ to $3$. The
relations 
between spinor-valued and normal (complex) tetrads are given by
\begin{displaymath}    
        e^{AA'}{}_{\mu} := e^{\al}{}_{\mu} \; \si_{\al}{}^{AA'}
        \qquad \mbox{and} \qquad  
        e^{\al}{}_{\mu} = - e^{AA'}{}_{\mu} \; \si^{\al}{}_{AA'},
\end{displaymath}   
where $\al$, $\be$ are flat indices running from $0$ to $3$. 
Flat indices are pulled up and down with the Minkownski metric
$\et_{\al \be} = \et^{\al \be} = diag(-1,1,1,1)$. 
The 
$\si^{\al}{}_{AA'}$ are the {\it Infeld van der Waerden symbols}, 
defined by \boldmath 
\begin{eqnarray*}    
        \mbox{\unboldmath $\si^{\al}{}_{AA'}$} & := &  \si_{\al} \\
\end{eqnarray*}
$\si_0$ \unboldmath
being  $- 1 / \sqrt{2}$ times the unit $2 \ti 2$ matrix whereas the
other 
\boldmath $\si$'s \unboldmath are $1 / \sqrt{2}$ times the Pauli
matrices. 
The outward normal spinor on a hypersurface 
described by the spatial components of the tetrad, $e^{AA'}{}_{i}$, 
$i = 1, 2, 3$, is defined by
\begin{displaymath}
  n_{AA'} \; n^{AA'} = 1 \qquad \mbox{and} \qquad n_{AA'} \;
  e^{AA'}{}_{i} = 0
\end{displaymath}
and is a function of the spatial components $e^{AA'}{}_{i}$ alone, given
by
\begin{displaymath}    
   n^{AA'} = \frac{i}{3 \sqrt{h}} \; \ve^{ijk} \; e^{AB'}{}_{i} \;
e_{BB'j} 
        \; e^{BA'}{}_{k}
\end{displaymath}
Using the relations for $n^{AA'}$ and  the 
properties of the Infeld van der Waerden symbols \cite{WB83}
one gets the following useful identities
\begin{eqnarray*}
e_{AA' i} \; e^{AB'}{}_{j} & = & - \frac{1}{2} h_{ij}
\epsilon_{A'}{}^{B'} 
- i \; \sqrt{h} \; \varepsilon_{ijk} \; n_{AA'} \;
e^{AB'k} \\
e_{AA' i} \; e^{BA'}{}_{j} & = & - \frac{1}{2} h_{ij} \epsilon_{A}{}^{B} 
+ i \; \sqrt{h} \; \varepsilon_{ijk} \; n_{AA'} \;
e^{BA'k} \\
n^{AD'} \; e_{BD'}{}^{i} & = & - n_{BD'} \; e^{AD'i}
\end{eqnarray*}
Spinor indices are contracted by the means of $\ep_{AB}$ od $\ep^{AB}$
according to 
\begin{displaymath}    
   \xi^{A} = \xi_{B} \; \ep^{AB} \qquad \xi_{A} = \xi^{B} \; \ep_{BA} 
\end{displaymath}
$\ep_{AB}$ is antisymmetric in its indices and obeys 
\begin{displaymath}   
        \ep^{AB} \, \ep_{BC} = \ep^{A}{}_{C} = - \de_{C}{}^{A} \qquad
        \ep^{AB} \, \ep_{CB} = \ep_{C}{}^{A} =  \de_{C}{}^{A}
\end{displaymath}
Analogous relations hold for $\ep_{A'B'}$.
From the definition of $n^{AA'}$ it follows that
\begin{displaymath}    
  n_{AA'} \; n^{AB'} = \frac{1}{2} \; \ep_{A'}{}^{B'}
\end{displaymath}
The one-component spinors used, like $\ps^{A}{}_{\mu}$, are taken to be 
Grassmann valued which directly leads  to the identity
\begin{equation}    
  \ps^{\ti A'}{}_{[\mu} \;  \ps^{\ti}_{A' \nu]} = 0
 \label{anticom}
\end{equation}

From (\ref{3.9}) one gets the useful formulas
\begin{eqnarray}  
    \ve^{nmp} \; D^{H}{}_{A'rn} \; D^{A'}{}_{Cmi} & = & - \frac{2
i}{\sqrt{h}} 
        \; n^{H}{}_{D'} \; e^{D'}{}_{Ci} \, \de^{p}{}_{r}
        \label{6.2} \\
    \ve^{irs} \; D^{DA'}_{ji} \; D^{C}{}_{A'kr} & = &  - \ve_{jkp} \;
h^{ps} \; \ep^{CD} 
        - \frac{2i}{\sqrt{h}} \; h_{kj} \; n_{A'}{}^{C} \; e^{DA's}
   \label{4.27}
\end{eqnarray}

%%%%%%%%%%%%%%%%%%%%%%%%%%%%%%%%%%%%%%%%%%%%%%%%%%%%%%%

\end{document}